\documentclass[11pt,twoside]{article}

\def\eps@scaling{.95}
\def\epsscale#1{\gdef\eps@scaling{#1}}

\def\insertplot#1#2#3#4#5#6#7{
\vskip 10pt\nobreak\hbox to \hsize{\hss\dimen0=#3in\hbox to #6\dimen0{%
\dimen0=#2in\vbox to #6\dimen0{\vss
\special{ps: plotfile #1}
\special{ps::[end]
  PGPLOT restore
}
}\hss}\hss}\vskip 10pt}

\usepackage{asp2006}
\usepackage{epsf}
\usepackage{psfig}
\usepackage{lscape}

\begin{document}
\title{An Assessment of HR Diagram Constraints on Ages and Age Spreads in Star-Forming Regions and Young Clusters}   %%% Fill in title
\author{Lynne A. Hillenbrand}   %%% Fill in author names
\affil{California Institute of Technology}    %%% Fill in author affiliations
\author{Amber Bauermeister} %\footnote{current affiliation UC Berkeley}}   %%% Fill in author names
\affil{California Institute of Technology and UC Berkeley}    %%% Fill in author affiliations
\author{Russel J. White}   %%% Fill in author names
\affil{University of Alabama - Huntsville}    %%% Fill in author affiliations

\begin{abstract} %%% Abstract to run on from here.
Pre-main sequence evolutionary theory is not
well-calibrated to observations.  With care, the observed quantities can be
converted into effective temperature and luminosity (i.e. the
Hertzsprung-Russell diagram) which the theoretical calculations also
predict as a function of stellar mass and age.  For a sample of nearby
young stellar clusters and associations
ranging in age from $<$1 Myr to $>$100  Myr, we have
tested the loci of luminosity as a function of effective temperature
against various sets of predicted pre-main sequence isochrones.  As we found
in Hillenbrand \& White (2004) which tested stellar masses, here for
the stellar ages there are two conclusions: some evolutionary calculations 
fare better than others in reproducing the empirical sequences, and systematic
differences between all pre-main sequence evolutionary calculations
and the data are apparent.  We also simulate hypothetical clusters 
of varying star formation history and compare the resulting HR diagram
predictions to observed clusters.  Our efforts are directed towards
quantitative assessment of {\it apparent} luminosity spreads 
in star forming regions and young clusters, which are often erroneously 
interpreted as {\it true} luminosity spreads indicative of {\it true} age spreads.
\end{abstract}

%%% MAIN BODY OF TEXT GOES HERE. CONSULT "INSTRUCTIONS FOR AUTHORS USING
%%% LATEX2E MARKUP", SECTIONS 2.3-2.6 FOR HELP WITH EQUATIONS, FIGURES,
%%% AND TABLES.

%\section{}   %%% Top level section head (remove "%" symbol)
%\subsection{}   %%% Second level section head (remove "%" symbol)
%\subsubsection{}   %%% Lowest level section head (remove "%" symbol)
%\section*{}    %%% Unnumbered top level section head (remove "%" symbol)
%\subsection*{}   %%% Unnumbered second level section head (remove "%" symbol)

\section{Introduction}
Stars form from giant molecular clouds which become unstable to 
fragmentation and subsequent collapse of dense cores.  Two main theories of
star formation suggest different timescales for this process.  
Ambipolar diffusion (e.g. Shu, Adams, Lizano, 1987) is a quasistatic process 
that can occur over a range of time scales from just a few million years up 
to perhaps ten million years.  Turbulent dissipation (e.g. Elmegreen 2000)
occurs within a much shorter time frame, essentially the dynamical
or crossing time which is only several hundred thousand ranging up 
to a million years, or so, for typical clusters.
Accurately estimating the age and age spread of stars in recently
formed clusters is one direct means for observationally constraining
this formation timescale.

How can the ages of young stars be inferred from observations?
Dynamical time scales can be
derived based on the spatial distribution and velocity dispersions of young
stars in star forming regions, as advocated by Tan et al. 2006
(arguing for slow star formation) and Hartmann et al. 2001 (arguing
for  rapid star formation).
Nuclear burning time scales, such as lithium depletion, 
can be compared to the theoretical physics of this process
as discussed by e.g. Palla et al. 2005 and Jeffries et al. 2005.
Stellar structure and atmosphere theory, i.e. the classical HR diagram,
is a standard tool for inference of physical parameters of stars
having all ages, and has been used in practice by many authors.

It is this last method that is discussed here since the stars 
suitable for study via the first two methods are only a subset of the sample
for which estimates of effective temperature and luminosity are now available
in the literature. The HR diagram has many shortcomings, elucidated
below.  However, making use of the tools one has and recognizing their
limitations is better than both alternatives: not making progress at all,
on the one hand, and, on the other,
overstating our understanding of pre-main sequence
evolution in young clusters due to a lack of attention to caveats 
and limitations of employed methodology.

\section{The Data}

The axes of the HR diagram, log $L/L_\odot$ and log $T_{eff}$, remain
difficult to determine with high precision for star forming regions.
Such quantities are derived from two sets of observations.  First, 
typically low resolution optical 
(usually in the V or I bands) or near-infrared (usually in the J or K bands)
spectroscopy measures photospheric emission and can be used for
spectral typing and thus, in combination with a gravity/metallicity 
dependent temperature scale, for temperature estimation.  Second, 
optical or near-infrared photometry is compared with intrinsic colors 
estimated from the spectral type, and used to infer foreground extinction.
The appropriate bolometric correction is then applied to reddening 
corrected photometry and a stellar luminosity is thus derived. Seemingly
straightforward, this process of course suffers many challenges in practice.

A recent example of the observational complications is provided by comparison 
of data from an HST wide-field ACS survey of the Orion Nebula Cluster
(Robberto et al. 2006) 
with older ground based data (Hillenbrand 1997) which shows considerably larger 
photometric scatter.  This is no doubt due to a combination of the large 
dynamic range in source brightness, the crowding of point sources, and 
the bright and
spatially variable nebular background, all of which can bias photometric data 
having relatively larger point spread function.  High data quality 
that can counteract the many observational challenges presented when working
in regions of recent star formation is a primary 
consideration in the quest for high fidelity age estimates from HR diagrams.

\begin{figure}[t]
{
\plottwo{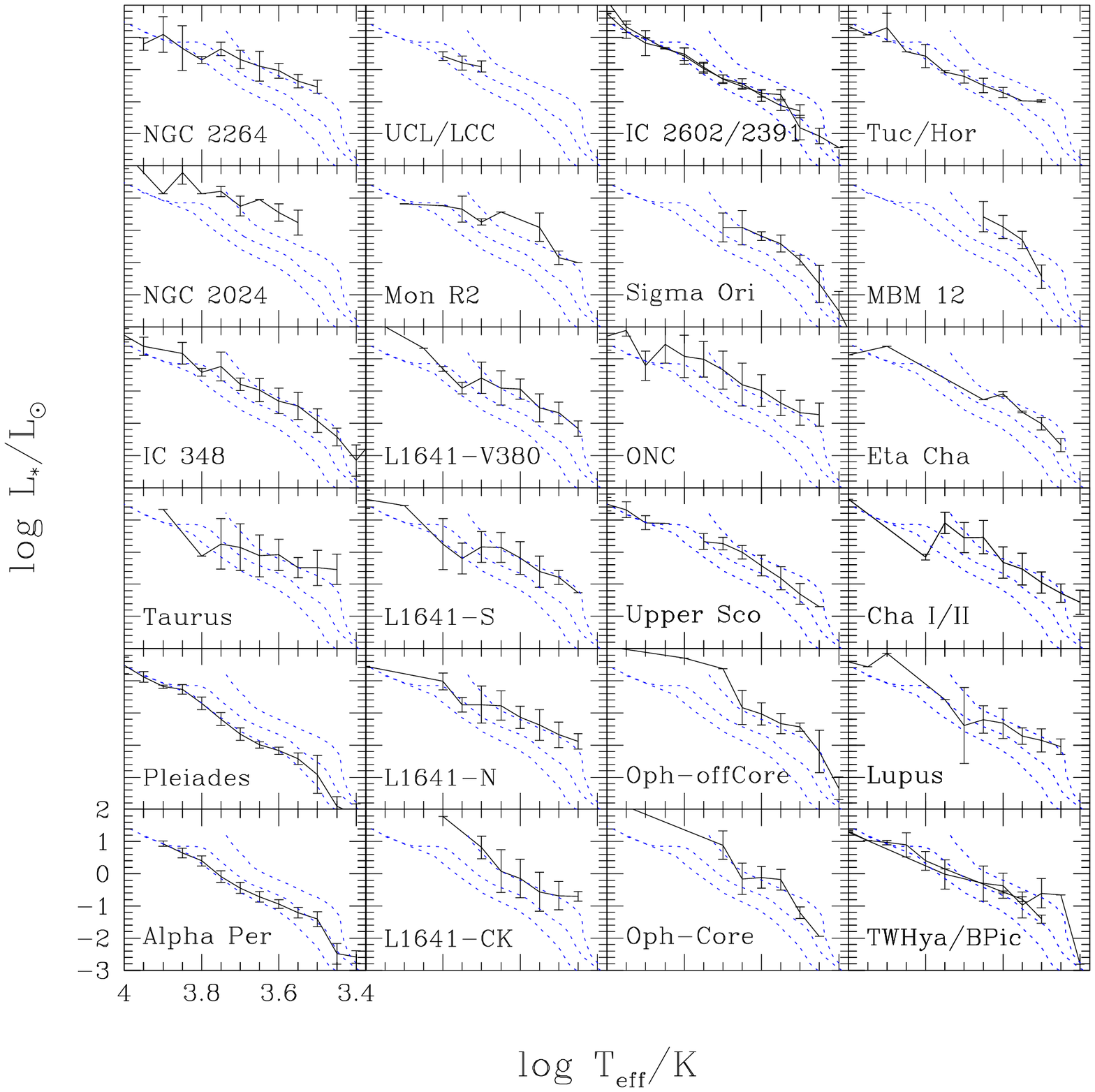}{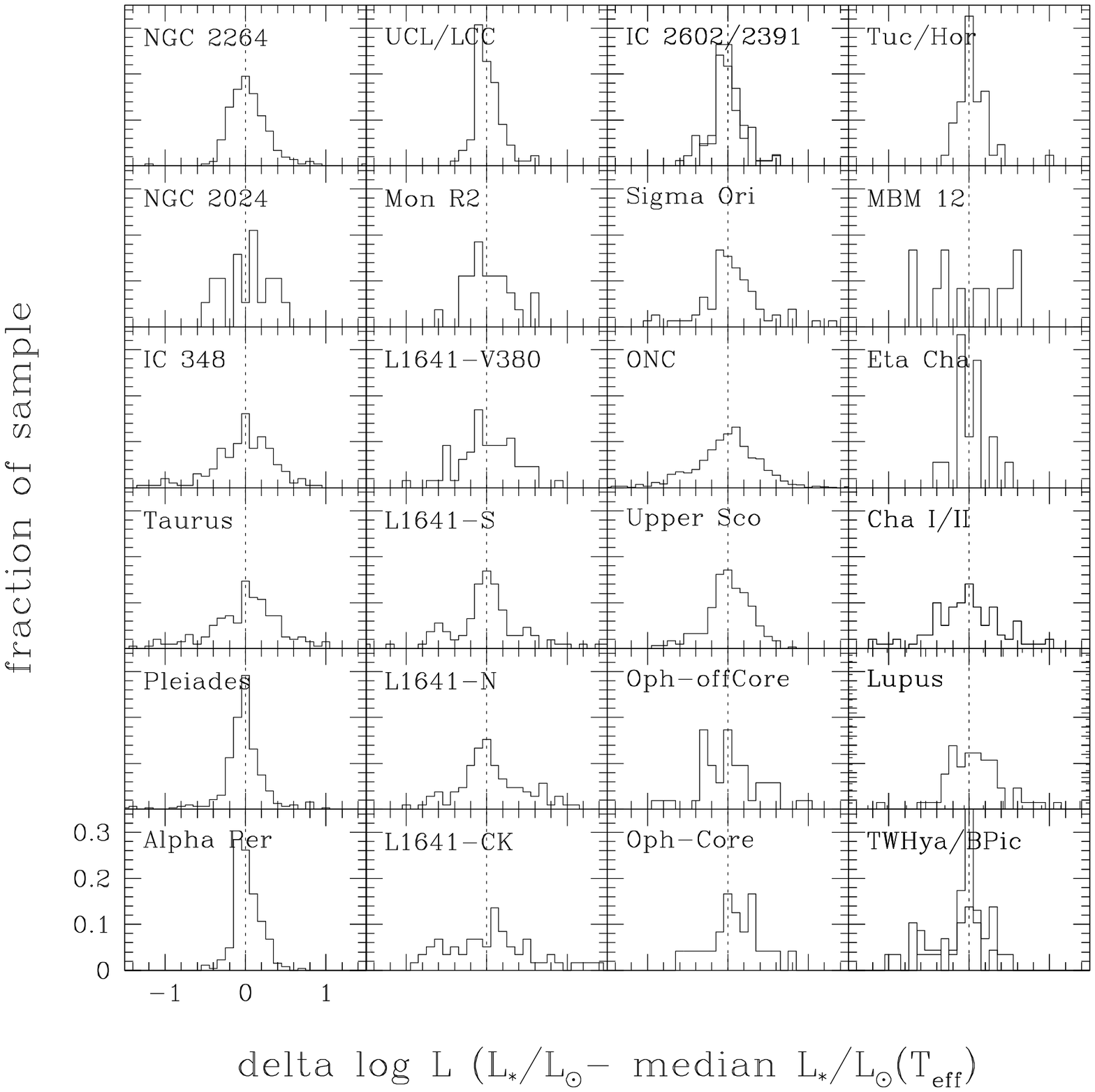}
%\epsscale{0.6}
%\plotone{hrds.meanl.dm98.ps}
%\newline
%\plotone{delmeanl.ps}
%\epsscale{1.0}
}
\caption{Representations of young star cluster luminosity spreads.
{\it Left panel}: Median and 1-sigma luminosity as a function of effective 
temperature, shown for spectral types cooler than A0 
(masses $<$3 M$_\odot$). Comparison of the empirical isochrones (solid lines)
is made to the 1, 10, and 100 Myr constant age sequences from 
D'Antona \& Mazzitelli (1997/1998; dotted lines).
{\it Right panel}: Luminosity dispersion calculated for individual data points
around the median luminosity at each effective temperature from the left panel.
These histograms collapse the two dimensional information of the HR diagram 
into one dimensional distributions, thus absorbing any trends in luminosity
dispersion with mass.
}
\label{fig:meanl}
\end{figure}

Further astrophysical, rather than mere observational, complications 
for young star cluster photometry include: random effects of
photometric variability, systematic effects of activity/disks, and systematic
effects of
spatially unresolved binaries.  Each is an interesting area of study
in its own right, but here we lump all such astrophysical 
effects into the observational error terms.  This results in errors in 
effective temperatures and luminosities that are larger than the formal errors
in these quantities one would estimate by simple error propogation
from the observed quantities (spectral types and photometry).

Without, for the moment, consideration of the above observational and 
astrophysical error terms and their effects on young star
temperatures and luminosities, what do we find 
when we assemble the HR diagrams for all well populated 
and well studied recently star forming regions and young
open clusters in the solar vicinity?  As we discuss in detail in a forthcoming
paper, there is variety in the richness levels in the known populations, 
as well as in data quality. Nevertheless, a composite of such HR diagrams 
clearly illustrates an age progression among the clusters from $<$1 to 120 Myr,
this via the decrease in mean/median luminosity due to the contraction of 
individual cluster members towards the zero age main sequence.  Because of 
the large number of data points, the main loci of points are easier 
to see by considering the median luminosity as a 
function of spectral type and the dispersion about this median 
(Figure~\ref{fig:meanl}a)
or the detailed luminosity distribution about this median 
(Figure~\ref{fig:meanl}b).  

In what follows we discuss such representations of
the data relative to predictions for simulated clusters.
The main question we aim to address is: do observed luminosity spreads 
correspond to age spreads, or are they consistent with error distributions 
created from the combined observational and astrophysical contributions
to stellar luminosity error terms? 

\section{The Isochrones}

Before we begin our assessement, a pre-cursor question to be answered is:
do theoretical pre-main sequence evolutionary tracks correctly predict 
stellar ages? If so, which set amongst those available and having considerable 
discrepancy among them in their predictions of effective temperature 
and luminosity at a given mass and age, should we believe?  

\begin{figure}[t]
\plottwo{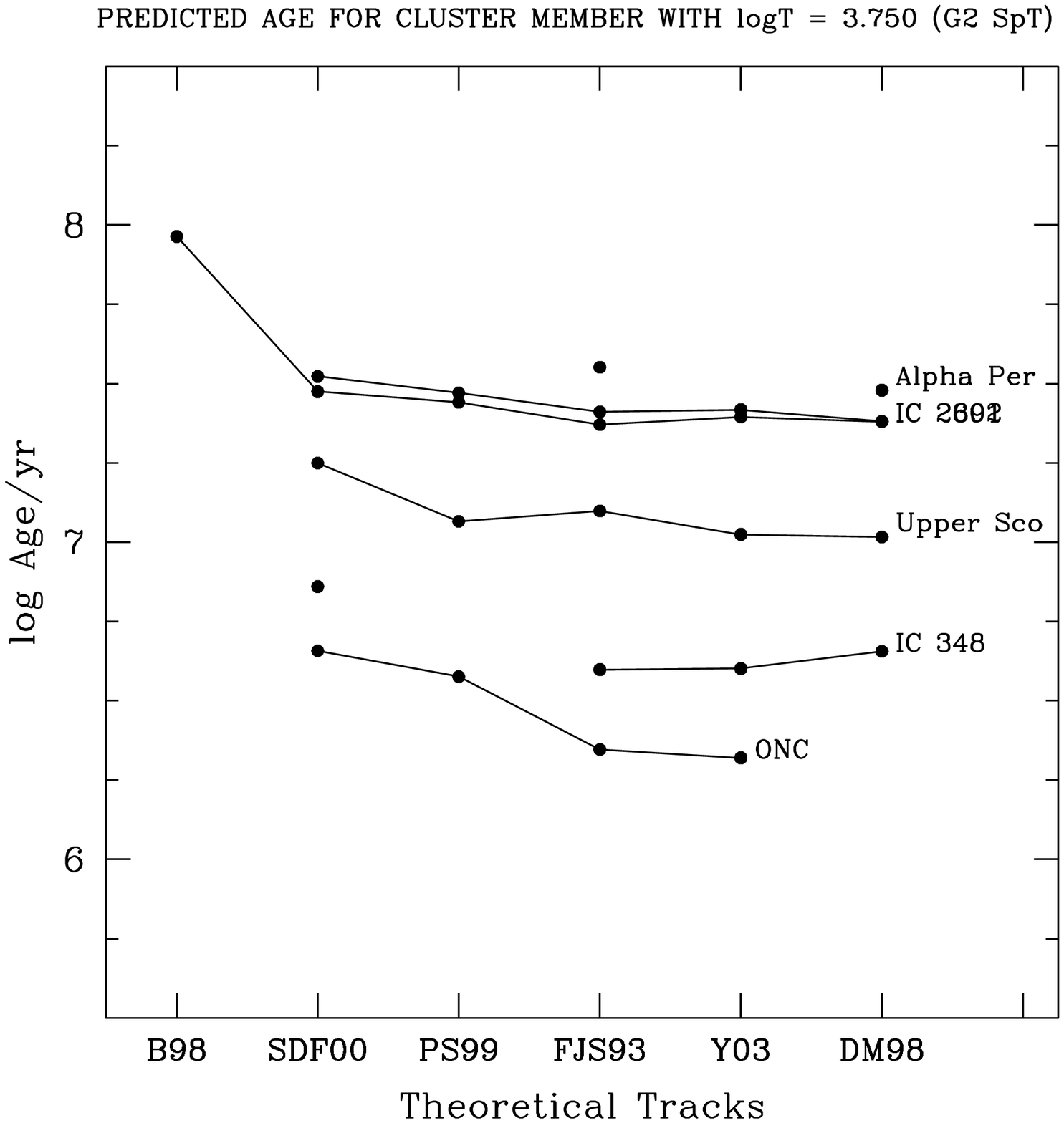}{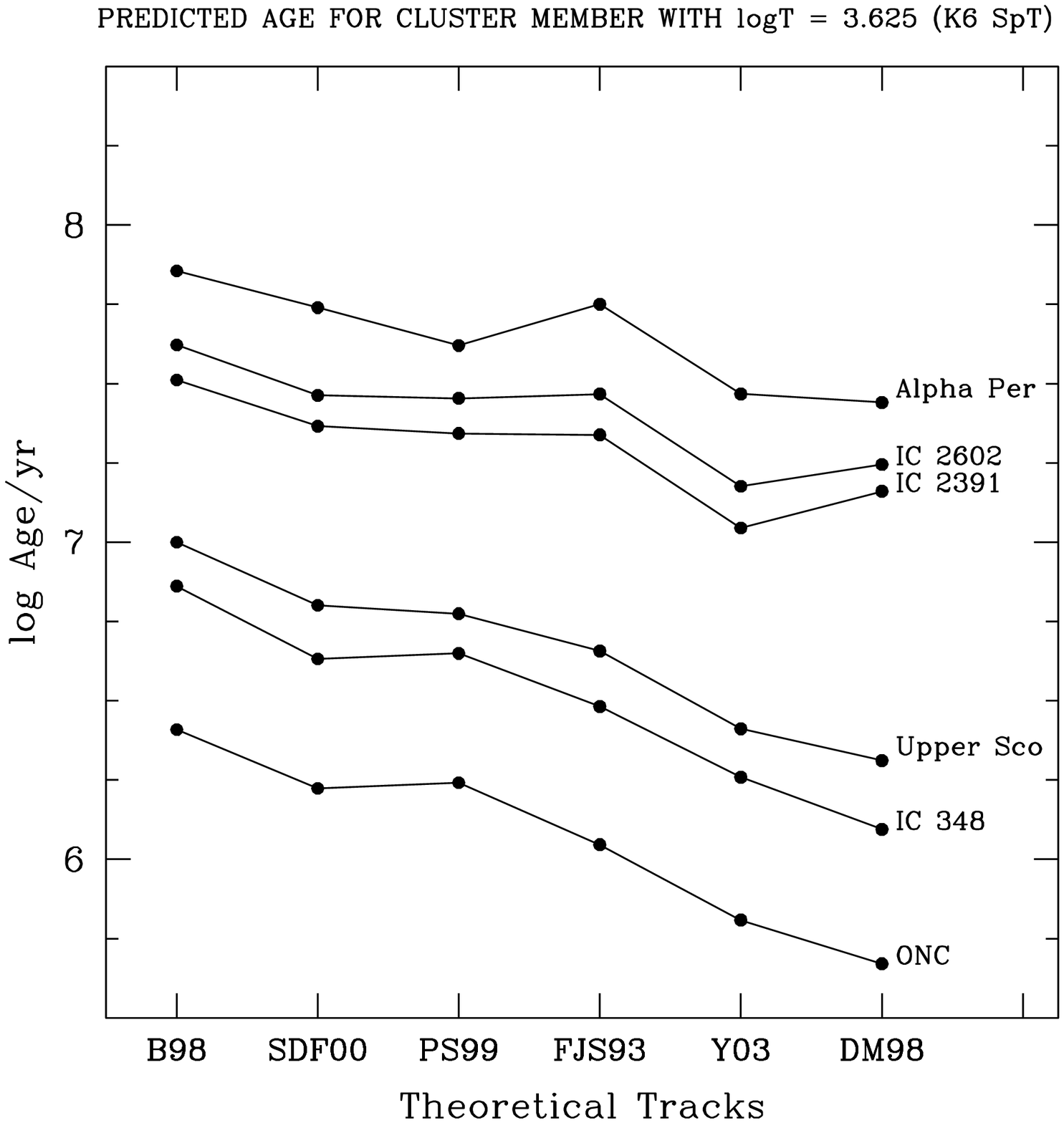}
\caption{Systematic trends in predicted stellar ages for G2 (left panel)
and K6 (right panel) stars having the median luminosity of the indicated 
clusters at those spectral types. Two effects are apparent.  
First is variation in ages predicted by different sets of
pre-main sequence evolutionary tracks for the same value of log $L/L_\odot$
and log $T_{eff}$.  Towards older ages especially, the G2 star predictions  
are relatively flat indicating that the models are fairly consistent with 
one another; the K6 star predictions, however, show considerably more 
variation between the models.
Second are the different ages predicted between the left and right panels 
for stars which lie along the same empirical isochrone (i.e. same age) 
but are of different spectral type (i.e. different mass).  Specifically, 
in Upper Sco the G2 stars have predicted ages somewhat older than 10 Myr,
fairly uniformly among the tracks, while the K6 stars are 2-10 Myr depending
on track choice.
}
\label{fig:systematic}
\end{figure}

There are at least 6 different groups with published pre-main sequence 
evolutionary calculations that have been widely circulated in machine-readable 
formats and that span a suitable range of stellar masses.
For these, listing only the most recent reference for each group
--  Swenson et al. 1994 (S93); D'Antona \& Mazzitelli 1997 
with 1998 electronic-only update (DM98); 
Siess et al. 2000 (S00); Baraffe et al. 1998 (B98) with Chabrier et
al. 2000; Palla \& Stahler 1999 (PS99); and Yi et al. 2003 (Y$^2$) --
the differences in assumptions, included physics, and methods
are broadly outlined in Hillenbrand \& White (2004) which assessed the
consistency of these models with available dynamical mass estimates.  
Here, we focus on the age predictions of these same models.

We can assess the systematic trends
between the various sets of tracks by considering the predictions for some
fiducial stars of given temperature and luminosity.  Figure~\ref{fig:systematic}
compares the ages inferred for young solar-type and low-mass members 
of several young clusters as modelled with the 6 sets of tracks.  
For the sub-solar mass stars, systematic effects between the tracks
are observed at the level of 0.75 dex at the youngest ages;
cluster age estimates are, therefore, strongly dependent on which set of 
pre-main sequence evolutionary theory is adopted.  For the solar-mass stars
the agreement is better, particularly towards older pre-main sequence ages.

Putting the systematics between the tracks, aside, for all tracks, 
the higher mass stars are predicted to be
older than the lower mass stars in the same clusters.  This effect 
often has been ascribed to the influence of the ``stellar birthline" 
(Stahler 1983; Hartmann, Cassen, \& Kenyon 1997).  However, 
comparing the left and right panels of Figure~\ref{fig:systematic}  
reveals that the observed age-with-mass trend persists longer than the
influence of birthline effects is expected to last.
%At the youngest ages no independent measures of stellar age are available,
%though several techniques such as pulsations and Li depletion trends
%show promise. At older ages there are independent checks that suggest 
%isochronal ages are systematically too young for pre-main sequence ages
%$>$20 Myr.

\section{Comparison of Simulated and Empirical Isochrones}

\begin{figure}[t]
%           #1: PostScript file name (PGPLOT /VPS file)
%           #2: figure height (in)
%           #3: figure width (in)
%           #4: lower left corner x (inches from llc of paper)
%           #5: lower left corner y (inches from llc of paper)
%           #6: desired magnification
%           #7: 1 => original in landscape mode, 0 => portrait
\insertplot{trend.epsi}{6.0}{3.0}{-0.5}{3.}{0.35}{1}
\caption{Change in ``slope" of the simulated HR diagram,
simply $\delta$ log $L/L_\odot$ / $\delta$ log $T_{eff}$ 
calculated in four bins of width 0.05 dex from log $T_{eff}= 3.55-3.70$,  
as a function of the binary fraction for a simulated 5 Myr old cluster.  
Each line represents a different pre-main sequence evolutionary model, 
as labelled.
} 
\label{fig:slope}
\end{figure}

\begin{figure}[t]
%\epsscale{0.7}
\plottwo{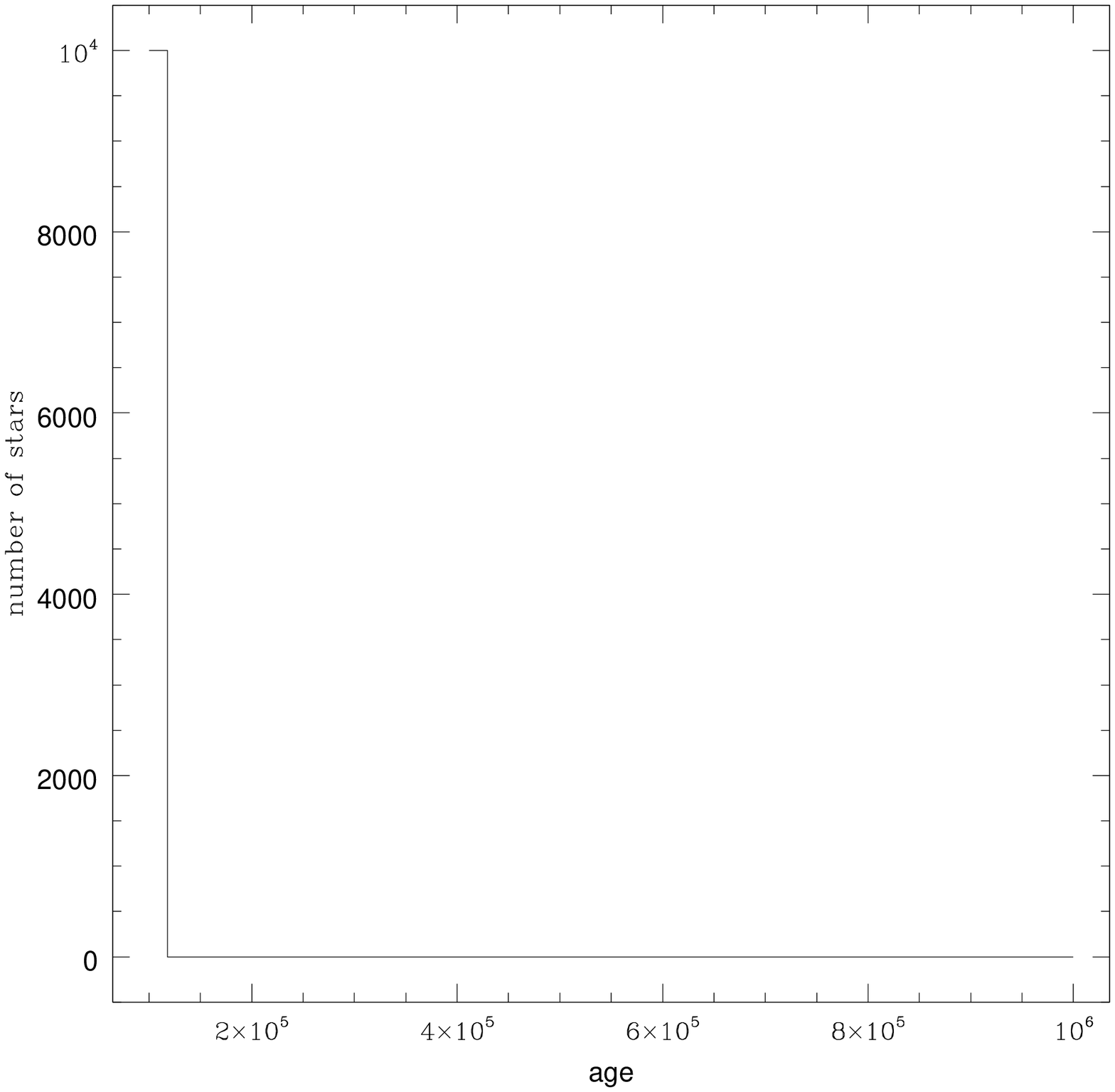}{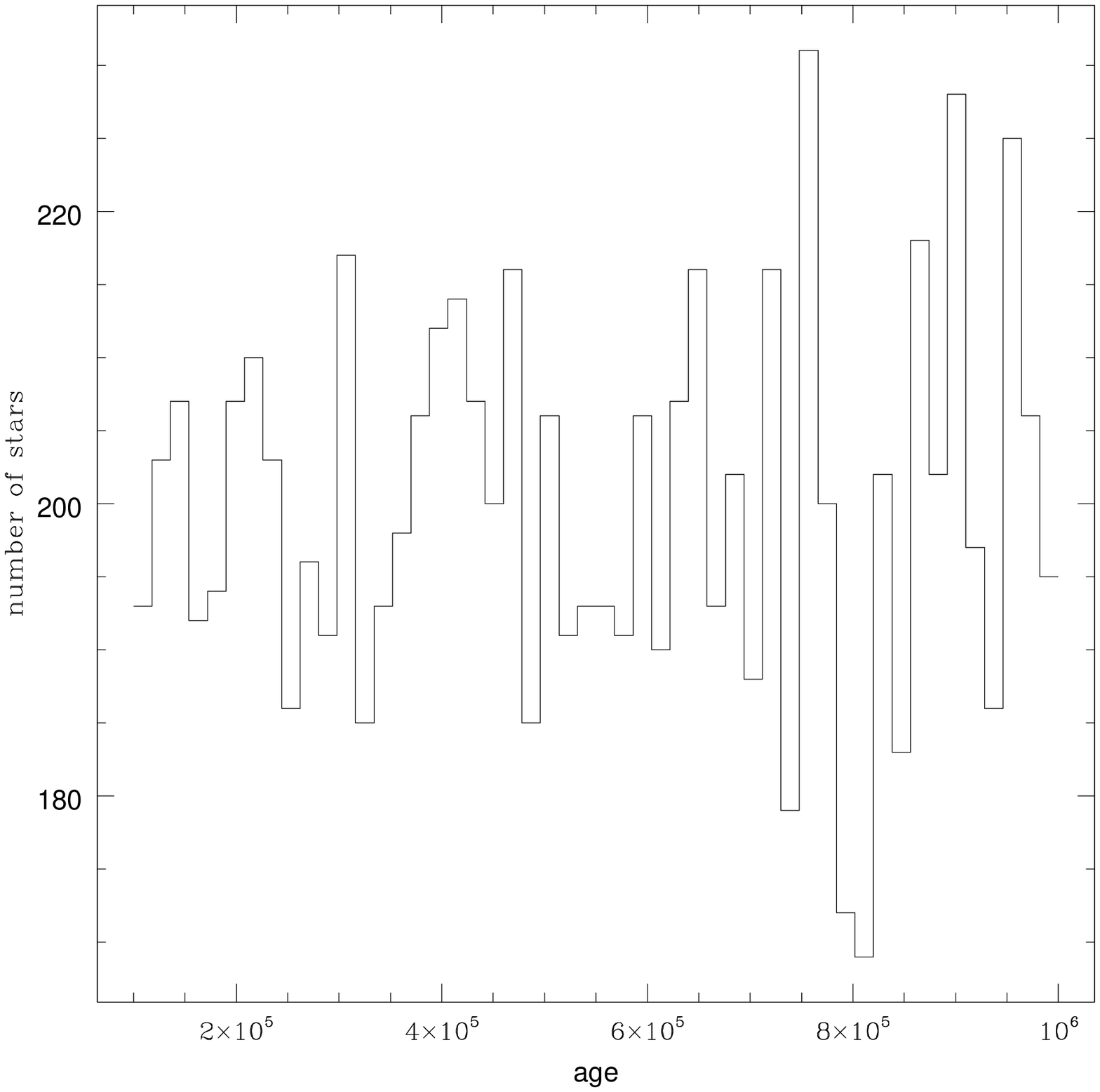}
\newline
\plottwo{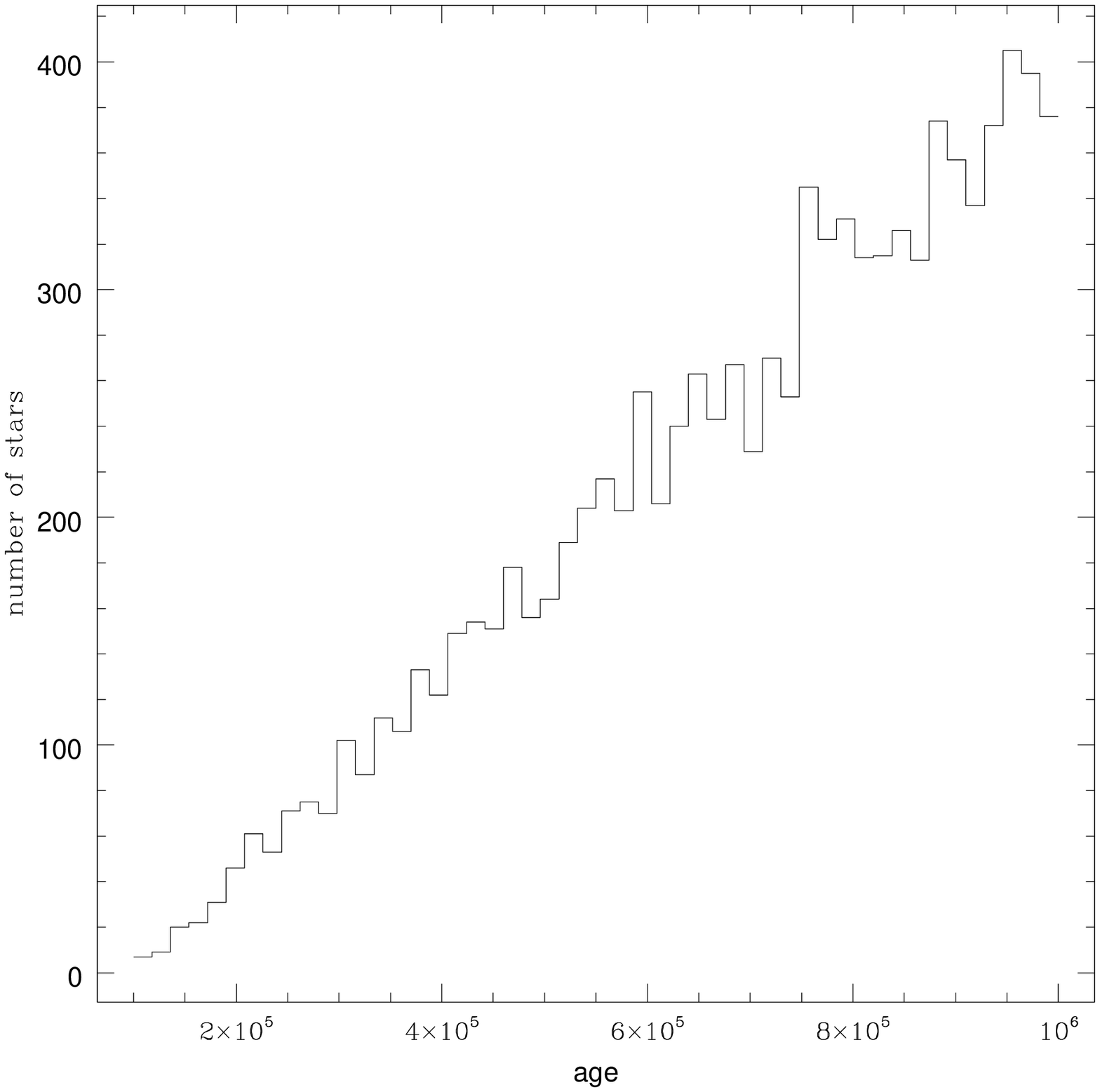}{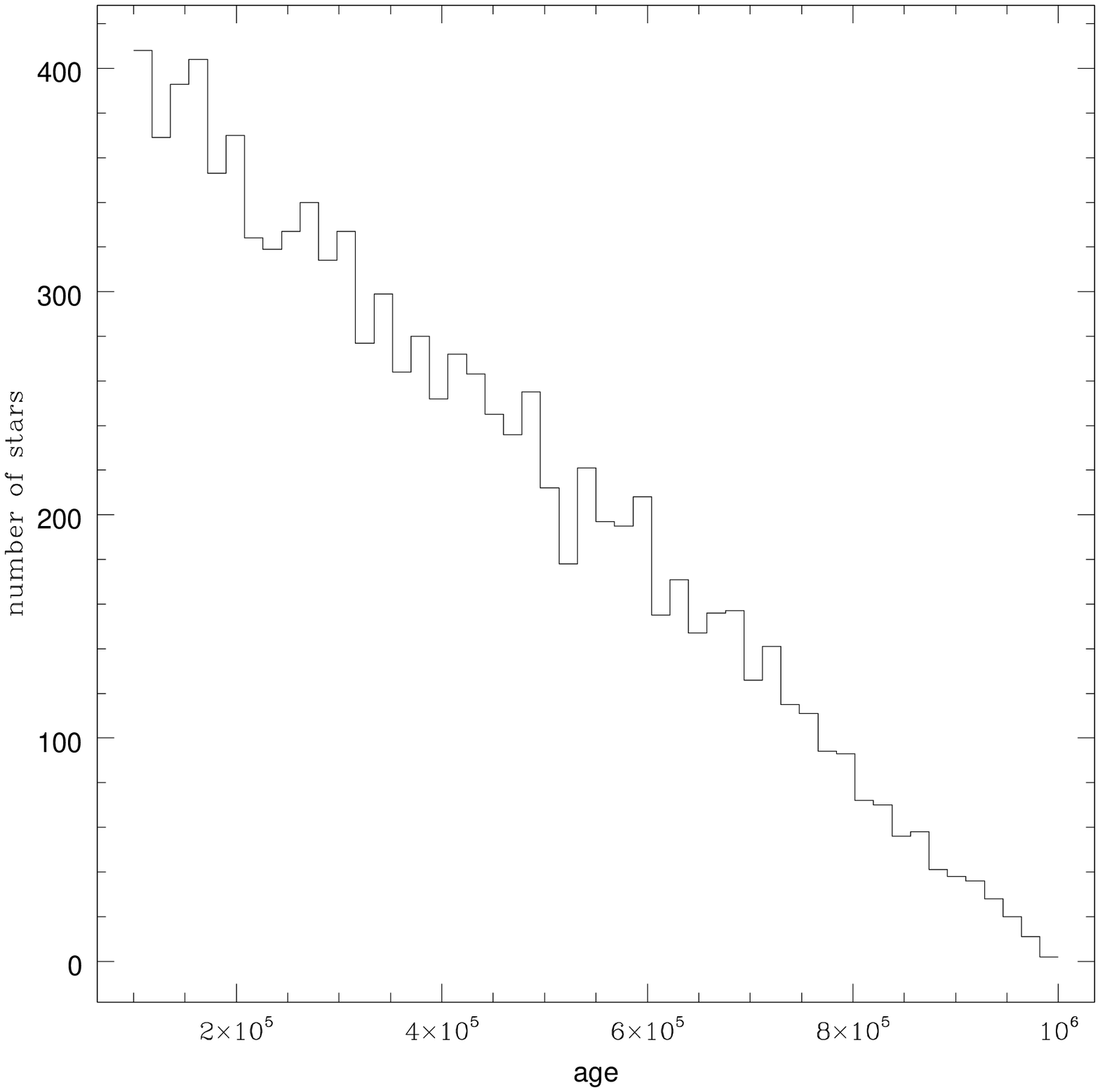}
%\epsscale{1.0}      
\caption{Simulated star formation histories: burst, constant, 
and linearly increasing/decreasing shown as number of stars formed 
vs time in years. Scatter represents numerical noise for a simulation
consisting of 10,000 stars.
}
\label{fig:agehist}
\end{figure}

In order to assess empirical ages and age spreads derived from HR diagrams,
we have created a suite of simulations that probe our ability to distinguish
true age spreads from observational and astrophysical noise.  We use
three diagnostics:  
\begin{itemize}
\item
the slope of the HR diagram, that is, 
[$\delta$(log $L/L_\odot$)/ $\delta$ (log $T_{eff}$/K)] 
calculated from the median luminosity as a function of effective temperature;
\item
the dispersion ($\sigma$) of the individual stellar luminosities measured around
the median luminosity at the same effective temperature;
\item
the detailed luminosity distribution
around the median luminosity at the appropriate effective temperature;
\end{itemize}

Roughly speaking, the clusters of Figure~\ref{fig:meanl}a follow
empirically a linear relation in log luminosity versus log temperature, at least
over the temperature range illustrated.  In Figure~\ref{fig:slope} we show
that the linear slope calculated for simulated HR diagrams of
given age not only varies between theoretical tracks, but is a strong 
function of the binary fraction.  For an assumed age of 5 Myr appropriate to
the Upper Sco association, one might conclude from Figure~\ref{fig:slope} that 
in comparison to the ``observed" slope for this cluster of 7.26, 
the Siess et al. tracks provide the best description of the data 
for a binary fraction near unity, while the Yi et al. tracks are best 
if the binary fraction is closer to zero; the D'Antona \& Mazzitelli tracks
might be appropriate for intermediate binary fractions.

Distinctions between young clusters are seen not only in these slopes, 
which represent the averaged empirical isochrone, but also in the detailed 
luminosity distribution about the median with effective 
temperature, which potentially represent age true age dispersions.
As was illustrated in Figure~\ref{fig:meanl}b, while some regions 
are reasonably well-described by gaussian luminosity distributions 
about the median value, suggesting that their luminosity spreads 
are consistent with errors, other regions show
a distinct step-like progression in the luminosity distribution towards
higher luminosities with a sharper fall-off in the distribution towards 
lower luminosities.  This form is most consistent with the expectations 
for a binary influence on the luminosity distribution, as illustrated below.

To investigate in more detail the consistency of observed luminosity spreads
with true age spreads, we employ a monte carlo methodology to populate 
theoretical pre-main sequence evolutionary tracks.  In our illustration of the
technique here, we adopt the Siess et al. tracks as our fiducial 
set.  Included in our simulations are a mass distribution (default assumption
is standard Miller-Scalo IMF) and a multiplicity fraction 
(default assumption is 40\% with secondaries drawn either from the IMF 
or from a simple functional form in $q = m_2/m_1$). 

\begin{figure}[t]
\plottwo{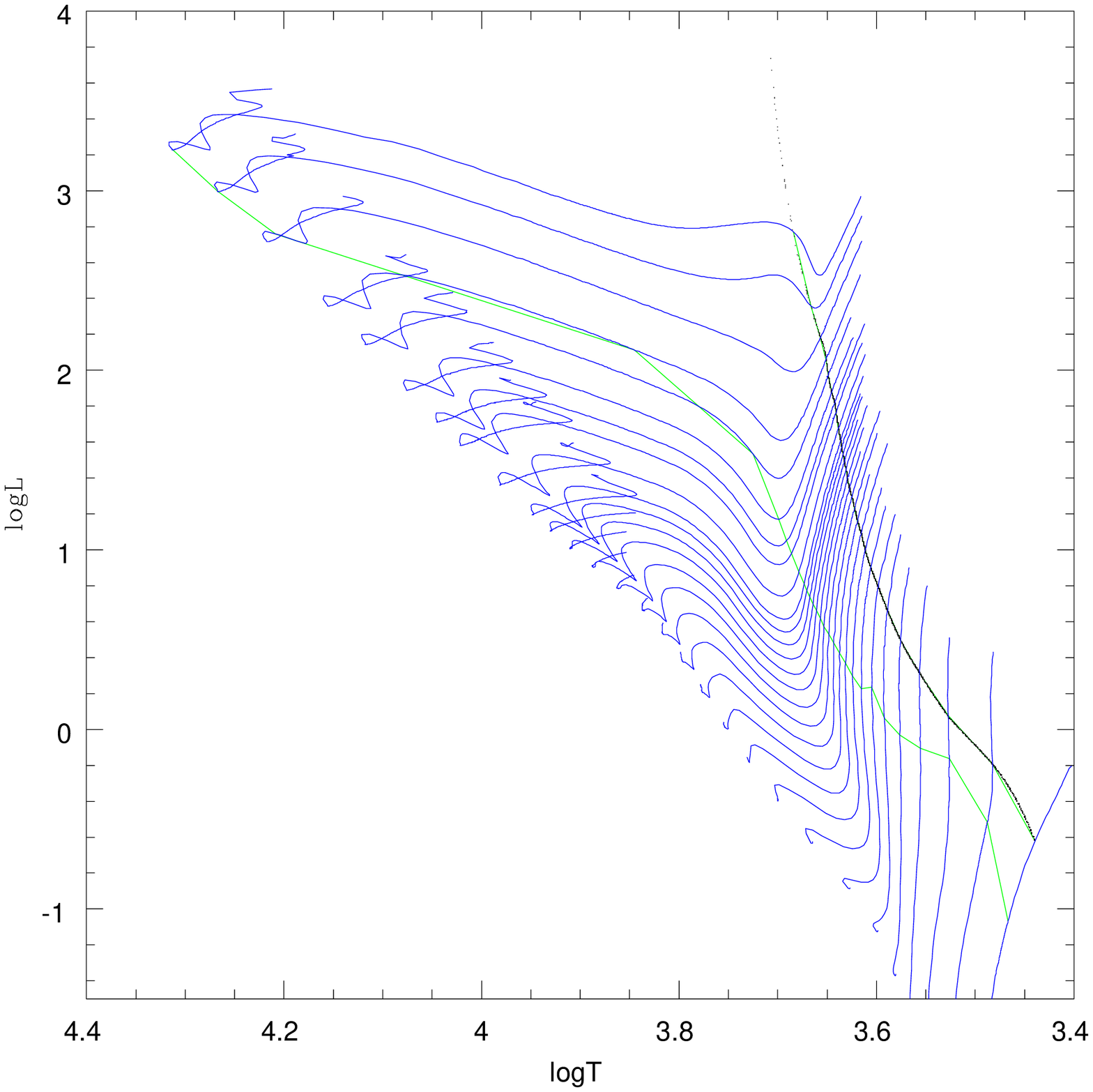}{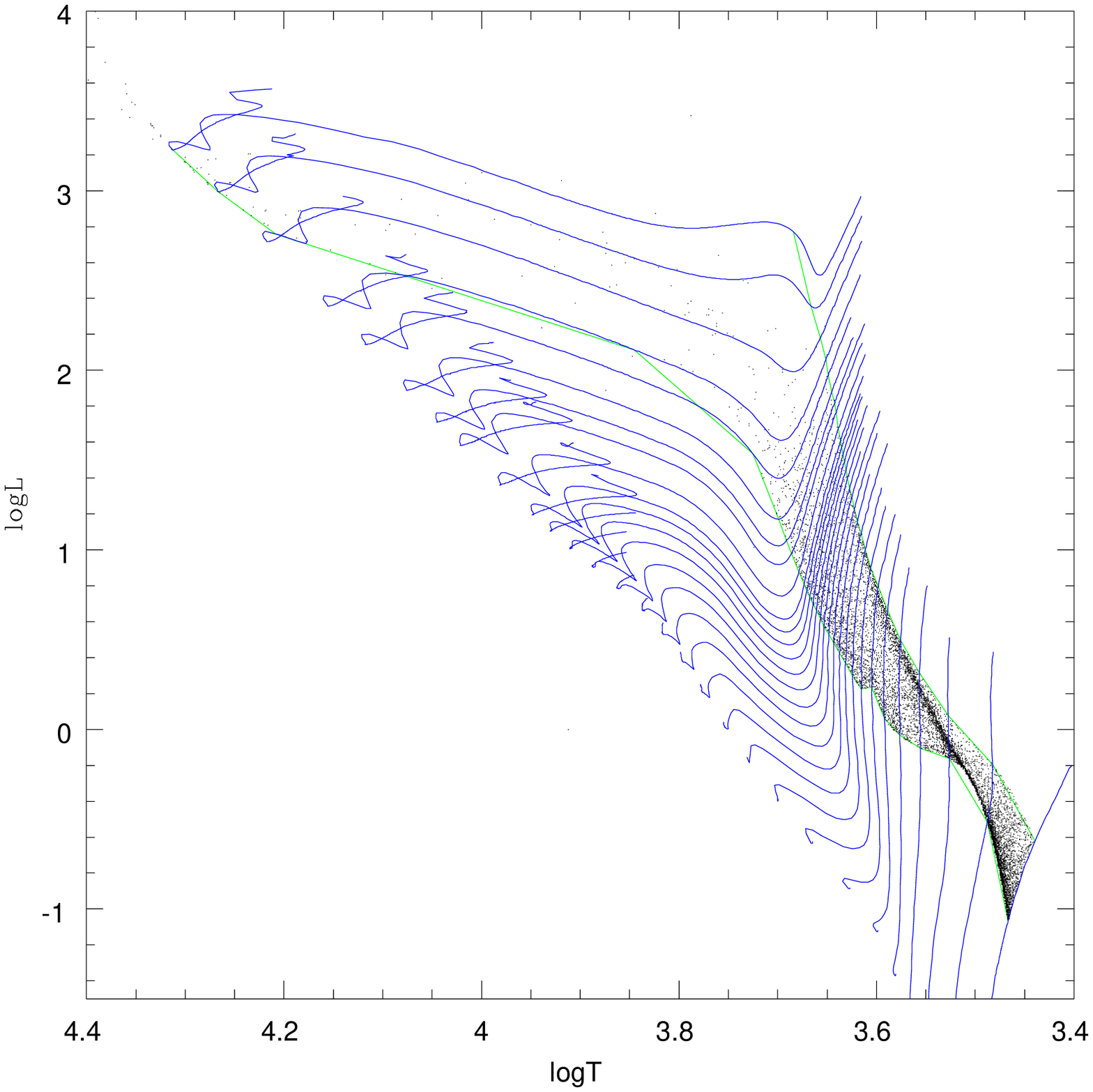}
\newline
\plottwo{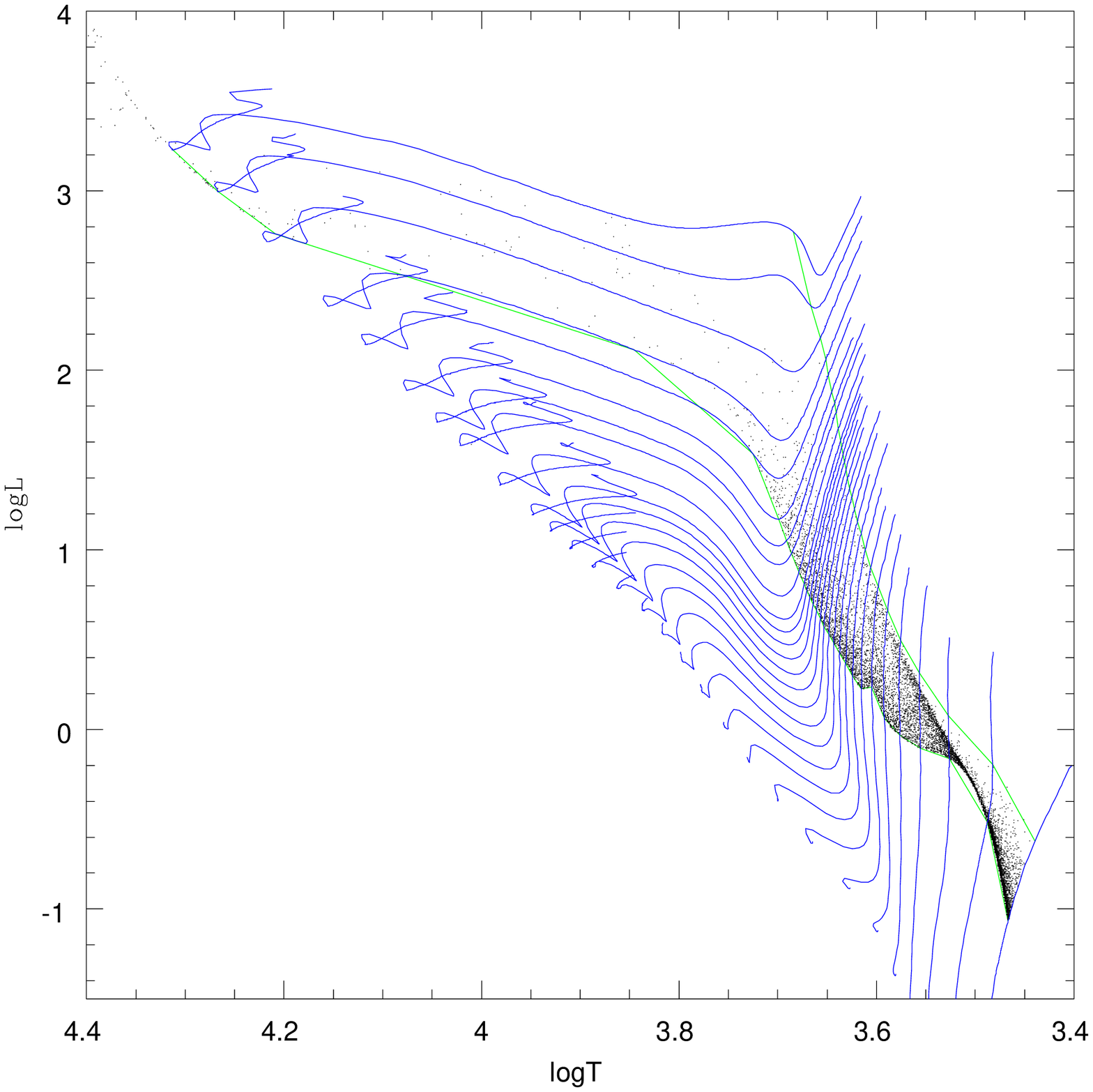}{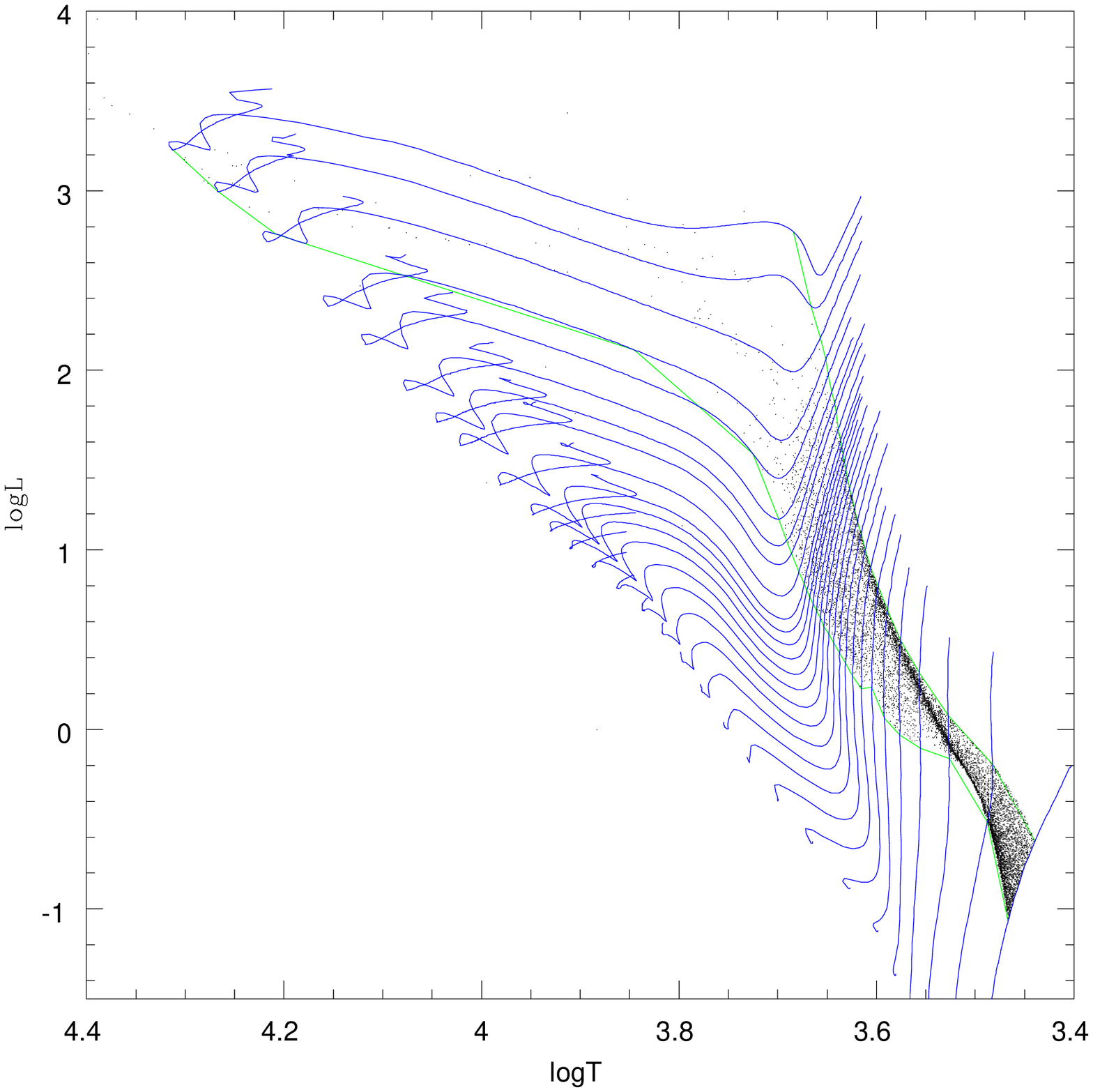}
\caption{Simulated HR diagrams in standard units
for the burst, constant, and linearly
increasing/decreasing star formation histories of Figure~\ref{fig:agehist}, 
using the Siess et al. (2000) tracks.  Mass tracks and isochrones at the 
start and end of the simulated age range are shown,
along with 10,000 simulated points.
}
\label{fig:hrdsim}
\end{figure}

Our main goal is to understand what can be inferred about star formation
histories from empirical data converted into an HR diagram.  Thus we have 
simulated various renditions of the sequence of star formation with
time in the observed young clusters.  We consider burst (no age spread)
scenarios as well as a constant rate of star formation with time and
linear or exponentially increasing and decreasing functions, as illustrated
in Figure~\ref{fig:agehist}. Example results for these star formation
histories are shown in the simulated but error-free HR diagrams of 
Figure~\ref{fig:hrdsim}. 
Finally, we add observational errors (uniform or gaussian) 
as the last step before creation of the HR diagram or of a histogram 
of the distribution of luminosity about the median luminosity run with
effective temperature.  

\begin{figure}
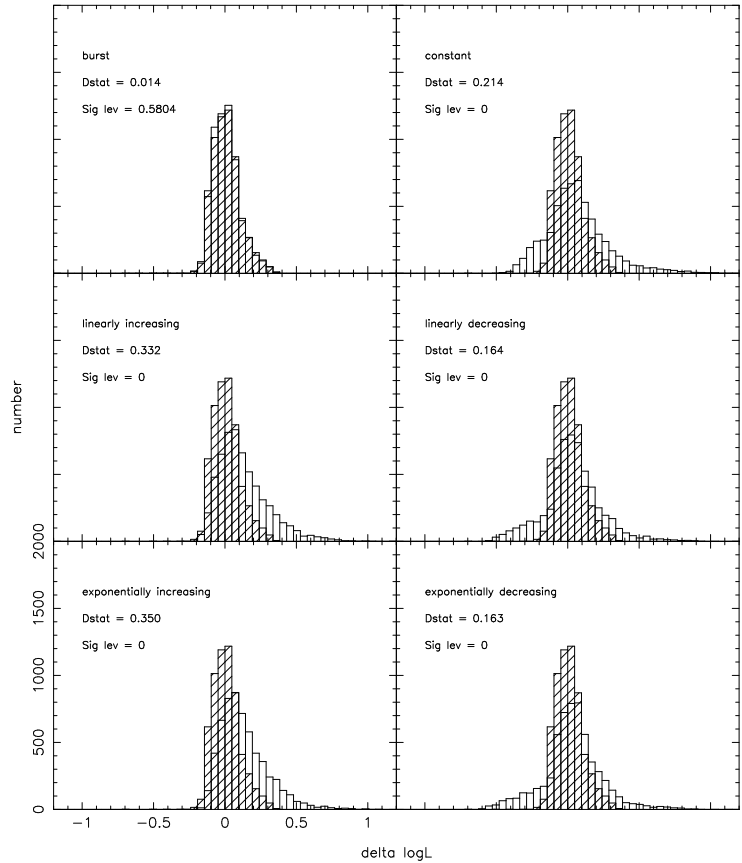

%\plottwo{../cluster.ages/dLerr_halfm.ps}{../cluster.ages/dLerr_twom.ps}
%\newline
%\plotone{../cluster.ages/dLerr_tenm.ps}
%\insertplot{../cluster.ages/amber.fig5.epsi}{7.5}{3.0}{0}{3.}{0.5}{1}
\insertplot{amber.fig5.epsi}{6.75}{3.0}{-0.5}{2.}{0.6}{1}
\caption{Simulated luminosity spreads (solar units), 
including effects of true age spread
plus observational error (open histograms) compared to burst 
or no age spread scenario, with error (hatched histogram, same in each panel). 
The KS ``D statistic" representing the
maximum difference between the cumulative distributions 
and its significance level are given.  Upper left panel represents two
realizations of the same star formation scenario and thus illustrates the
magnitude of numerical noise.  In all other panels the KS significance 
is better than one percent, suggesting we can distinguish the scenarios.
}
\label{fig:sfsim}
\end{figure}

\section{The Findings}

\begin{figure}
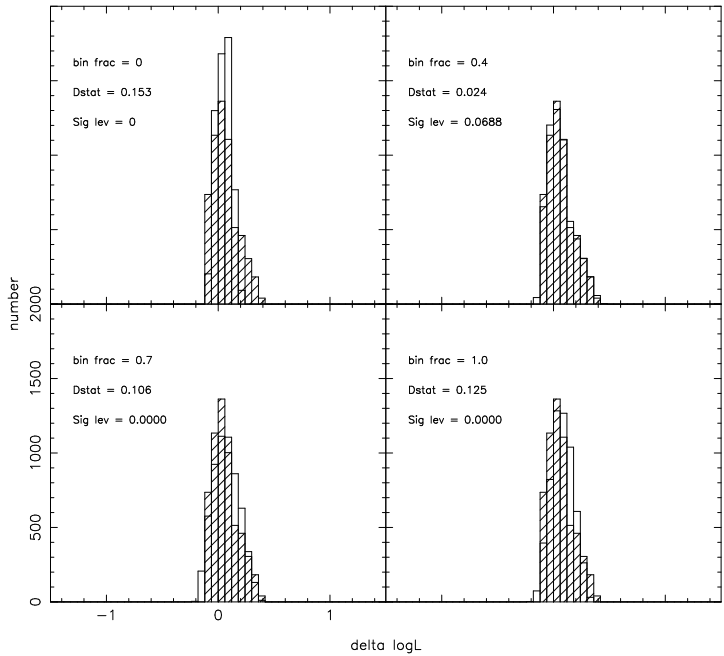

\insertplot{amber.fig7.epsi}{5.5}{3.0}{-1.5}{2.}{0.6}{1}
\caption{Simulated luminosity spreads (solar units)
for a 40\% binary fraction population 
from a ``burst" (no age spread) star formation episode 2 Myr ago including
observational errors (hatched histogram, which is the same in all panels).
This is compared to populations having binary fraction 0, 70, and 100\% 
all with a constant rate of star formation and 10\% age spread around
a mean age of 2 Myr (open histograms).  The upper right panel 
is thus not just two realizations of the same binary fraction;
the KS test finds that these cases still have a 7\% chance of being the same. 
In all other panels the KS significance is better than one
percent, arguing that binary properties can be detected against the
background of small age spreads.
}
\label{fig:binsim}
\end{figure}

Having presented the data and our simulation methods, we proceed now with
some illustrative simulation results. In Figures~\ref{fig:sfsim} 
and ~\ref{fig:binsim} we show the distribution of individual simulated
luminosities about the median luminosity appropriate to the relevant 
simulated effective temperature.
%XXXX 
%In practice this is done by calculating the average age of the central median
%luminosity points, creating an isochrone with fine sampling in log T at that
%average age, and finally calculating the $\delta$log L of the points
%from the isochrone.
%XXXX 

In the case of Figure~\ref{fig:sfsim} 
we are testing various star formation scenarios containing an 80\% 
age spread (that  is, for a cluster of median  age 1 Myr, 
the youngest stars are only 0.2 Myr old and the oldest are 1.8 Myr old)
against a 0\% age spread or a ``burst" star formation scenario. 
The ``core" in the delta-log$L$ 
distribution is due to the assumed error distribution (here 0.1 dex), 
the high log$L$ tail is due to the assumed binaries (40\% fraction), 
and the broad wings are indicative of the inserted age spread.  
To quantify what is visible by eye, we employ the Kolomogorov-Smirnov (KS)
test which produces the probability that two distributions are drawn from
the same parent distribution.  Here, except in the case of the burst
scenario, the KS test rejects that the input age spread produces the same
luminosity spread as the burst case.

In Figure~\ref{fig:binsim} we look at the ability of the simulations 
to distinguish multiplicity fraction. Specifically, we test a 40\% binary,
coeval population fiducial sample against populations with a mere 10\% age 
spread and different binarity fractions.  The histograms indicate a narrow 
excess in the 0\% binaries panel relative to the fiducial, and a broader excess 
in the 70 and 100\% binaries panels.  Here, the KS test finds that 
the 0, 70, and 100\% binaries cases are rejected as being drawn from the same
population, while the two 40\% binary fractions have reasonable chance of being
from the same parent.

\begin{figure}
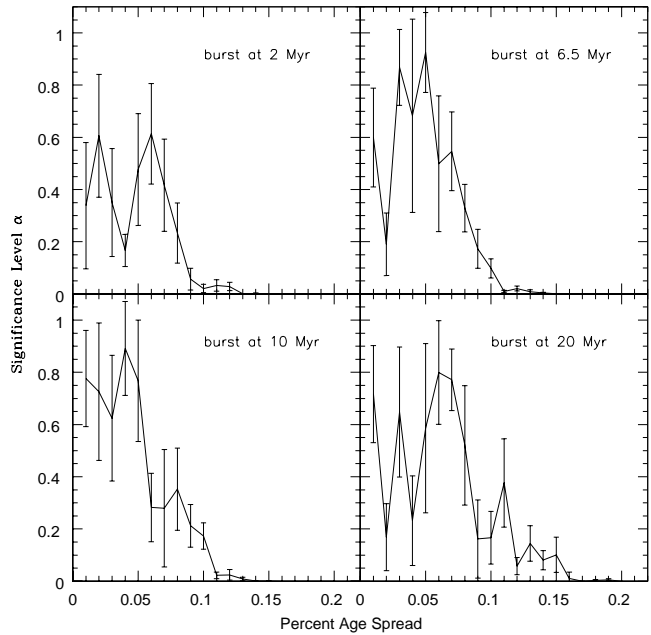

%\plotone{amber.fig9.epsi}
\insertplot{amber.fig9.epsi}{7}{3.0}{3.25}{2.25}{0.45}{0}
\caption{Results of KS tests. The decline towards zero indicates 
that an age spread of the indicated fraction along the abscissa 
{\it -- not percent, as incorrectly labelled --}
of the fiducial age (2, 6.5, 10, or 20 Myr depending on the panel) 
is distinguishable from a burst model having an error distribution
with maximum amplitude 0.1 mag.  Vertical bars indicate the dispersion 
amongst 100 simulations having each age spread.
}
\label{fig:ks}
\end{figure}

The question at hand is whether we can distinguish age spreads from either
the details of the star formation history or from the mulitiplicity effects.
To test this we have run a large number of simulations at median ages of
2, 6.5, 10, and 20 Myr and calculated the KS significance 
of the input fractional age spread compared to a zero
percent age spread.  Based on the decline of the KS statistic, where 
small implies distinguishable distributions, we conclude from 
Figure~\ref{fig:ks}) that when observational errors are modest 
($\pm$10\% on stellar luminosities), 
age spreads larger than $\sim$10-15\% can be distinguished.  

How good do the empirical luminosities really need to be?  More or less, 
the above scaling is roughly correct. At 1-10 Myr absolute ages, 
30\% luminosity errors mean that at best 30\% age spreads can be distinguished 
from 0\% age apreads; however, 30\% age spreads can not be distinguished 
from 5, 10, 20, 40, or 50\% age spreads. 
Our current work is to quantify more usefully our conclusions regarding
observational errors and cluster age spreads.

\section{Summary and Implications}
In summary, we offer a few simple cautions regarding stellar models and
physical parameters derived from classical HR diagrams.  First, 
pre-main sequence evolutionary tracks:
1) vary between theory groups;
2) under-predict stellar masses by 30-50\%  
(as assessed in White and Hillenbrand, 2004);
3) under-predict low-mass stellar ages by 30-100\%; 
and 4) over-predict high-mass stellar ages by 20-100\%.
These findings imply large systematic uncertainties in:
cluster initial mass functions, cluster age distributions, and hence
star formation histories in molecular clouds as well as
disk and angular momentum evolutionary time scales.

From our study of young star luminosities we have found useful 
diagnostics in the HR diagram slope, the median luminosity 
as a function of effective temperature, and the simple
dispersion as well as the detailed shape of the normalized luminosity 
function about the median run with effective temperature.  
From our simulations we conclude that 
observed HR diagram slopes can inform track choice modulo binarity
and that KS tests of luminosity distribution about median can distinguish:
multiplicity fraction, star formation history, and even true age spreads
given small enough observational errors.

Finally, we conclude based on our (in)ability to distinguish signal
from observational and astrophysical noise in the HR diagram, 
that at present there is only marginal, i.e. no strong, 
evidence for moderate age spreads in recently star forming regions 
and young open clusters (see also, Hartmann 2001).  
These findings are consistent with the decades old
-- but unheeded -- warnings on young cluster HR diagrams that were 
issued by Larson (1972) on their utility for understanding 
pre-main sequence evolution, and by Mercer-Smith et al. (1984) on the
interpretation of luminosity spreads as true age spreads.

%\acknowledgements %%% Text of acknowledgements runs on after this command.

%%% THE BIBLIOGRAPHY
%%%
%%% CONSULT SECTION 3 OF "INSTRUCTIONS FOR AUTHORS" FOR HOW TO USE NATBIB.
%%% AUTHORS ARE ENCOURAGED TO USE EITHER THE "THEBIBLIOGRAPY" ENVIRONMENT
%%% BY UNCOMMENTING (DELETING THE "%" SYMBOL) THE COMMANDS BELOW, OR BY
%%% USING THE BIBTEX ENVIRONMENT. TO FIND OUT WHICH IS APPLICABLE TO YOUR
%%% CONTRIBUTION, CONSULT THE VOLUME EDITORS FOR YOUR PROCEEDINGS.
%%%


\begin{thebibliography}{}
\bibitem[]{}
Baraffe, I., Chabrier, G., Allard, F., \& Hauschildt, P. H. 1998, AA, 337, 403 (
B98)
\bibitem[]{}
D'Antona, F., \& Mazzitelli, I. 1997, in Cool stars in Clusters and Associations
, ed. R. Pallavicini, \& G. Micela, Mem. S. A. It., 68, 807 (DM97)
\bibitem[]{}
Elmegreen, B.G. 2000, ApJ, 530, 277
\bibitem[]{}
Hartmann, L., 2001, AJ 121, 1030
\bibitem[]{}
Hartmann, L., Ballesteros-Paredes, J. \& Bergin, E.A. 2001, ApJ, 562, 852
\bibitem[]{}
Hartmann, L. Cassen, P., \& Kenyon, S. J. 1997, ApJ, 475, 770
\bibitem[]{}
Hillenbrand, L.A., 1997, AJ, 113, 1733
\bibitem[]{}
Hillenbrand, L.A. \& White, R.J., 2004, ApJ, 604, 741
\bibitem[]{}
Jeffries, R.D., 2005, in ``Chemical Abundances and Mixing in Stars in the
Milky Way and its Satellites," eds L. Pasquini and S. Randich (Springer-Verlag),
in press (astro-ph/0411111)
\bibitem[]{}
Larson, R. 1972, MNRAS, 156, 437 
\bibitem[]{}
Mercer-Smith, J.A., Cameron, A.G.W., \& Epstein, R.I. 1984, ApJ, 279, 363
\bibitem[]{}
Palla, F., \& Stahler, S. W. 1999, ApJ, 525, 772 (PS99)
\bibitem[]{}
Palla, F., Randich,S., Flaccomio, E., \& Pallavicini, R., 2005, ApJ, 626, 49
\bibitem[]{}
Robberto, M., O'Dell, R.C., Hillenbrand, L.A., Simon, M. et al. 2005, AAS, 207, 1460 (abstract only)
\bibitem[]{}
Shu, F., Adams, F.C., \& Lizano, S. 1987, ARAA, 25, 23
\bibitem[]{}
Siess, L., Dufour, E., \& Forestini, M. 2000, AA, 358, 593 (S00)
\bibitem[]{}
Stahler, S.W. 1983, ApJ, 274, 822
\bibitem[]{}
Swenson, F.J., Faulkner, J., Rogers, F.J., \& Iglesias 1994, ApJ, 425, 286 (S93)
\bibitem[]{}
Tan, J.C., Krumholz, M.R., \& McKee, C.F., 2006, ApJ, 641, 121
\bibitem[]{}
Yi, S., Kim, Y.-C. \& Demarque, P. 2003, ApJS, 144, 259
%\bibitem[]{}
\end{thebibliography}
\end{document}